# The Negative Action Keldysh Spinors

A. Jourjine[1]

FG CTP
Dresden, Germany

## Abstract

We discuss quantization of Dirac spinors with action which is the negative of the standard action. Using the Keldysh QFT formulation, we show that the standard quantization of such *Keldysh spinors* results in the Hamiltonian bounded from above and in S-matrix for backwards in time scattering. The alternative quantization produces the standard Hamiltonian bounded from below and the standard S-matrix for forwards in time scattering. When the two masses are equal, the sum of their actions describes the standard Keldysh out-of-time-order correlators. When both Keldysh and Dirac spinors are present in a QFT, the standard Fock space is augmented by negative energy states, while its positive and the negative subspaces share the same vacuum state. In absence of gravity the Dirac and the Keldysh spinors do not interact and positive and negative energy states do not exchange quanta with non-zero energy. If Keldysh spinors exist in Nature, then the Universe would consist of bosonic and fermionic particles that are described by both positive and negative Fock energy states with a single vacuum state that cannot be crossed from either side. Such a scenario could explain the non-observation of dark energy/dark matter except via classical gravity.

---

[1] jourjine@pks.mpg.de

## 1. Introduction and Summary of Results.

Quantization of quantum field theories with fermionic degrees of freedom is well-known to be ambiguous, because of the anti-commuting nature of the fermionic creation and annihilation (c-a) operators. This is remarkably different from quantization of the bosonic fields, which is essentially unique, except for possible zero point energy contribution uncertainty. During the second quantization of spinor fields there are four possible ways to assign quantum c-a operators to obtain a free quantum field from its classical analog. Two of them result in the indeterminate Hamiltonian, which is clearly unphysical, because of the absence of the vacuum state, while the remaining two quantizations produce Hamiltonians differing by a sign. The former is bounded from below, while the latter from above. We will call them the s-quantization and the a-quantization, s- for the standard and a- for the alternative. The standard S-matrix uses s-quantization, while the a-quantization is considered unphysical. Here we take this point of view under a review. In fact we will prove below that if s-quantization generates forwards in time scattering S-matrix, then a-quantization generates backwards in time scattering S-matrix.

Recently a class of the so-called flavor spin QFTs was described, where spinors of some flavors contribute to the total action with the negative of the standard action [1, 2, 3]. They appear when fermionic fields are described by the fermionic differential forms. The appearance of the negative contributions in these theories is unavoidable, because it the consequence of the use of the Lorentzian metric. In such theories the signature of the kinematic flavor quadratic form is generically pseudo-unitary, for example, $(+,+,-,-)$ for four generations or $(+,+,-)$ for three generations, instead of the standard $(+,+,+)$ in the SM. Flavor spin theories are attractive because not all Yukawa mass matrices are allowed. The restrictions lead to constraints on the flavor mixing, resulting in the observed flavor mixing matrix textures of quarks and leptons without the need for additional field content. Another theory where the total action is the sum of the positive and the negative action parts is well-known as the Keldysh formulation of the finite temperature QFT. Below we will describe the exact correspondence between Keldysh QFT and the flavor spin theory. The quantization of negative action spinor fields was briefly discussed in [2]. Here we provide a more detailed analysis and point out an intriguing consequence of the existence of such quantum fields, namely the non-interaction of some quantum versions of the fields. This might have a relevance to the solution of the dark energy/matter puzzles.

Our main result is the proof that the s-quantization of the negative action spinors is identical to the a-quantization of the positive action spinors, while the a-quantization of the negative action spinors is identical to the s-quantization of the positive action spinors. We will prove that if $S_{s,a}(\pm A)$ are the S-matrices for bilinear in spinor fields spinor action $A$ for different quantizations and action signs and interaction Lagrangian $\mathcal{L}_I$

$$S_s(+A) = T\exp\left[-i\int_{-\infty}^{+\infty} d^4x (\mathcal{L}_I)^s\right], \qquad (1.1)$$

then
$$S_a(+A) = \tilde{T} \exp\left[+i\int_{-\infty}^{+\infty} d^4x (\mathcal{L}_I)^a\right], \tag{1.2}$$

and

$$S_a(-A) = S_s(+A), \tag{1.3}$$

$$S_s(-A) = S_a(+A), \tag{1.4}$$

where $\tilde{T}$ is the reversed time ordering.

For application to flavor spin QFT equation (1.3) is most significant. It signifies that under the a-quantization the negative action (Keldysh) spinors are indistinguishable from the standard spinors. This means that when a-quantized the theories with pseudo-unitary kinematic flavor quadratic form are well defined as a standard QFT and can be treated on equal footing with the standard Dirac spinor fields.

Equation (1.4) also has some interesting consequences. It says that the standard quantization of Keldysh spinors describes the field theory on the negative direction time branch of the Keldysh QFT. The s-quantized Keldysh spinors are simply the standard Dirac spinors propagating backwards in time. In addition, we will see that the sum of Dirac and Keldysh actions with the same mass is identical to a single Dirac field of the Keldysh formalism that propagates both forwards and backwards in time, depending on the time branch where it is located. Since forward and backward time direction fields do not interact in principle (in effect they are located in the absolute past/future of each other) from the positive time direction fields point of view they act as dark matter. The non-interaction should also apply to graviton exchange in quantum gravity. However, classical static gravitational fields play a special pre-quantization role for definition of single particle fermionic states. Therefore, indirect interaction of Keldysh and Dirac spinors is possible after all.

Another consequence of (1.4) is the possible explanation of missing 4[th] quark and 4[th] lepton generations in 4-generation flavor spin theory. It can explain why the three-generation mixing matrices for quarks and leptons that are the best phenomenological fit can be obtained by elimination of one row and one column from 4-dimensional versions of $V_{CKM}$ and $U_{PMNS}$ [1,3].

## 2. The Keldysh Spinors and their Quantization.

Let us now discuss the negative action spinors in more detail. The free field action for such spinors is not surprisingly given by

$$S_K = -\int d^4x\, \bar{\psi}\, (i\slashed{\partial} - m)\psi\,. \tag{2.1}$$

We shall call such fields the *Keldysh spinor* or *K-spinor fields*, because Keldysh QFT formalism plays an important role in understanding of their unusual nature. Obviously the equations of motion for both Dirac (*D*) and K fields are the same, while for the Hamiltonian we obtain $H_K = -H_D$.

The standard quantization produces operator $H_D$ bounded from below, the alternative from above. For $H_K$ the boundedness properties are reversed. Note, that the transition from the standard to the alternative quantization may be considered as a generalized symmetry, an involution acting on the action $I: S_{K,D} \to -S_{K,D} = S_{D,K}$. This is a transformation that does not involve fields or coordinates. It leaves the field variables $\psi(x)$ intact, but it reverses the sign of all momenta $\pi(x)$ and of the Hamiltonian

$$I: \psi(x) \to \psi(x), \qquad \pi(x) \to -\pi(x), \qquad H \to -H, \qquad I^2 = id. \tag{2.2}$$

Obviously the Poisson brackets and the Hamiltonian equations of motion do not change and in this sense it is a canonical transformation. If standard symmetries are described as transformations that leave action unchanged, in our case this condition is relaxed: only the invariance only of the equations of motion is required.

Proceeding further, we point out that positive and negative energy particles could well co-exist in a physical world, provided they do not interact. The argument that a theory that contains fermionic and bosonic fields with both positive and negative Hamiltonians is unphysical assumes that during interaction with positive energy particles the negative energy particles can acquire infinitely negative energy while the positive energy particles would acquire infinitely positive energy. Such a situation is definitely unphysical. However, this scenario is based on the condition that the two types of fermions can interact. If for some reason they cannot interact or interact only via zero energy quanta exchange, then the catastrophe cannot take place. When the two types of fields cannot interact, than it is perfectly acceptable to construct an extended Fock space from two Fock spaces with positive and negative energy states and with a single shared vacuum that cannot be crossed from below or from above. Note, that by interaction we mean exchange of quantum particles with positive energy. Interaction with a classical background field, for example with classical gravitational field is another topic that we will not discuss here.

Next, recall the expression for the evolution of an operator $A(t)$ from its initial state $A(t_i)$ to the final state $A(t_f)$ in the Heisenberg representation

$$A(t_f) = \exp(iH(t_f - t_i)) A(t_i) \exp-(iH(t_f - t_i)). \tag{2.3}$$

If we replace $H$ in with $-H$, we see that the evolution of $A(t)$ is equivalent to swapping $A(t_i)$ and $A(t_f)$ while keeping $(t_f - t_i)$ unchanged, i.e., to evolution with the original $H$ but backwards in time with positively increasing time lapse parameter. Obviously, the same argument also applies to the interaction representation. This obvious remark is actually one of the central points in our discussion. It implies that negative action theories are somehow related to positive action theories propagating backwards in time.

At this point it is helpful to begin using quantum field theory formalism, where time evolution in both directions enters on equal footing. Such an extended formalism has been described by Keldysh [4]. Since then it has been developed further considerably and applied to a large variety of non-equilibrium systems. For a review see [5, 6]. Although it is mostly used in the solid state physics and quantum statistics to describe non-equilibrium QFT at finite temperature, it can also be used at zero temperature. In fact Keldysh formalism in a certain sense is equivalent to the S-matrix formalism in QFT: the scattering amplitudes computed in one are equivalent to those computed in the other [7]. In a nutshell, a single Keldysh S-matrix describes two copies of the standard S-matrices, one for forwards in time scattering and one for backwards in time scattering.

In the Keldysh QFT one considers the evolution of Greens function as a the vacuum expectation value of $T_C$-product computed along the time path contour $C$ that extends from $-\infty$ to $+\infty$ for the positive time direction branch and then back to $-\infty$ on the negative time direction branch. One constructs Greens functions of the fields with time coordinates on both positive and the negative time branches. The definition of the two point Greens functions and their perturbation theory then can be considered from a unified point of view. The matrix of four Greens functions is assembled into a single one

$$G(x,y) = \begin{bmatrix} G^T(x,y) & G^<(x,y) \\ G^>(x,y) & G^{\tilde{T}}(x,y) \end{bmatrix}, \tag{2.4}$$

where, taking into the account that $t_-$ lies in the absolute future of $t_+$, we obtain for $x_\pm = (t_\pm, \vec{x})$

$$G^T(x,y) = -i\langle 0|T_C\psi(x_+)\bar{\psi}(y_+)|0\rangle, \quad G^T(p) = \frac{1}{p^2 + m^2 - i\varepsilon}, \tag{2.5a}$$

$$G^{\tilde{T}}(x,y) = -i\langle 0|T_C\psi(x_-)\bar{\psi}(y_-)|0\rangle, \quad G^{\tilde{T}}(p) = -\frac{1}{p^2 + m^2 + i\varepsilon}, \tag{2.5b}$$

$$G^>(x_+, y_-) = -i\langle 0|T_C\psi(x_+)\bar{\psi}(y_-)|0\rangle, \ G^>(p) = 2\pi i \delta(p^2 + m^2)\Theta(p^0), \tag{2.5c}$$

$$G^<(x_-, y_+) = -i\langle 0|T_C\psi(x_-)\bar{\psi}(y_+)|0\rangle, \ G^<(p) = -2\pi i \delta(p^2 + m^2)\Theta(p^0), \tag{2.5d}$$

and where $t_+$ is the point on the positive time direction path branch while $t_-$ is a point on the negative time direction path branch. $\tilde{T}$ is the T-product on the negative time path branch, that is with respect to negative time direction. The coupling constants acquire a minus sign on the negative time branch and symbolically can be assembled into

$$g \rightarrow \begin{bmatrix} g & 0 \\ 0 & -g \end{bmatrix}. \tag{2.6}$$

Using these generalized Greens functions, one can construct a well-defined interaction representation and perturbation expansion, where $G(x, y)$ enters as the main object.

The crucial point here is that the coupling matrix is diagonal. The fields on the different time branches do not and cannot interact [6, 7]. Since we can identify the negative Hamiltonian dynamics with propagation backwards in time, the same applies to the negative Hamiltonians. Keldysh and Dirac spinor fields do not interact on the quantum level.

The proof of (1.2 – 1.4) follows from explicit computations of Feynman propagators for Keldysh and Dirac spinors with s- or a-quantizations. Recall the standard mode expansion for Dirac fields under s-quantization

$$\psi_D(x) = \int \frac{d^3k}{(2\pi)^3} \frac{m}{k^0} \left( b_r(\vec{k}) \, u^r(\vec{k}) e^{-ikx} + d_r^+(\vec{k}) v^r(\vec{k}) e^{ikx} \right) \tag{2.7}$$

$$\overline{\psi}_D(x) = \int \frac{d^3k}{(2\pi)^3} \frac{m}{k^0} \left( b_r^+(\vec{k}) \overline{u}^r(\vec{k}) e^{ikx} + d_r(\vec{k}) \overline{v}^r(\vec{k}) e^{-ikx} \right), \tag{2.8}$$

where $k^0 = +\sqrt{\vec{k}^2 + m^2}$. The plane-wave solutions $u^r(\vec{k})$, $r=1,2$, for the positive and $v^r(\vec{k})$, $r=1,2$, for the negative energy satisfy $(\slashed{k} - m)u^r(\vec{k}) = (\slashed{k} + m)v^r(\vec{k}) = 0$ and are normalized in the standard way: $\overline{u}^p(\vec{k})u^q(\vec{k}) = -\overline{v}^p(\vec{k})v^q(\vec{k}) = \delta^{pq}$. Here $b_r^+(\vec{k})$, $b_s(\vec{k})$ are the anti-commuting c-a operators for the Dirac particles. The energy momentum operator $P^\mu$ is

$$P^\mu = \int \frac{d^3k}{(2\pi)^3} \frac{m}{k^0} k^\mu \left( b_r^+(\vec{k}) b_r(\vec{k}) + d_r^+(\vec{k}) d_r(\vec{k}) \right), \quad \langle 0 | P^\mu | 0 \rangle \geq 0. \tag{2.9}$$

The a-quantization of Dirac fields swaps c-a operators so that $P^\mu \rightarrow -P^\mu$. For Keldysh fields s- and a-quantizations are reversed with the same results.

The standard Feynman propagator is given by

$$S_F(x-y) = -i \langle 0 | T \psi(x) \overline{\psi}(y) | 0 \rangle = \int \frac{d^4k}{(2\pi)^4} \frac{(\slashed{k}+m)}{k^2 - m^2 + i\varepsilon} e^{-ik(x-y)}. \tag{2.10}$$

In perturbation theory it determines the forward in time S-matrix. The backwards in time two-point function turns out to be

$$\widetilde{S}_F(x-y) = -i \langle 0 | \widetilde{T} \psi(x) \overline{\psi}(y) | 0 \rangle = -\int \frac{d^4k}{(2\pi)^4} \frac{(\slashed{k}+m)}{k^2 - m^2 - i\varepsilon} e^{-ik(x-y)}, \tag{2.11}$$

Further, we obtain that

$$S_K(x) = \tilde{S}_F(x), \quad (2.12)$$

$$\tilde{S}_K(x) = S_F(x). \quad (2.13)$$

S-matrix is uniquely determined by n-point Wightman correlators and those are in turn perturbatively determined by two point functions via Wick's theorem. Now we can use the fact that for theories with bi-linear spinor interactions coupling constants change sign with the change of quantization

$$(g)_{s,a} \to -(g)_{a,s}, \quad (2.16)$$

and with this in mind we conclude that

$$S_a(-A) = S_s(+A), \quad (2.14)$$

$$S_s(-A) = S_a(+A). \quad (2.15)$$

This concludes the proof of equations (1.2 – 1.4).